\documentclass{elsart}

\usepackage{amsmath}%
\usepackage{amssymb}%
\usepackage{graphicx}
\usepackage{dcolumn}
\usepackage{bm}

\newcommand{\bi}[2]{\left(\begin{array}{c}#1\\#2\end{array}\right)}

\begin{document}

\begin{frontmatter}

\title{Hidden structure in the randomness of the prime number sequence?}

\author[a1,uc3m]{S. Ares\thanksref{saul}}
\author[a1,upco]{M. Castro\thanksref{mario}}
\address[a1]{Grupo Interdisciplinar de Sistemas Complejos (GISC, {\tt
http://gisc.uc3m.es}) and}
\address[uc3m]{Departamento de Matem\'aticas,
Universidad Carlos III de Madrid\\ Avenida de la Universidad 30, 28911
Legan\'es, Madrid, Spain.}
\address[upco]{Grupo de Din\'amica No Lineal (DNL),
Escuela T\'ecnica Superior de Ingenier\'{\i}a (ICAI),
Universidad Pontificia Comillas, 28015 Madrid, Spain.}
\thanks[saul]{saul@mpipks-dresden.mpg.de}
\thanks[mario]{marioc@upco.es}

\begin{abstract}
We report a rigorous theory to show
the origin of the unexpected periodic behavior seen in the
consecutive differences between prime numbers. We also check numerically our
findings to ensure that they hold for finite sequences of primes, 
that would eventually appear in applications. Finally,
our theory allows us to link with three different but important
topics: the Hardy-Littlewood conjecture,
the statistical mechanics of spin systems, and the
celebrated Sierpinski fractal.
\end{abstract}

\end{frontmatter}

\section{Introduction}
\begin{figure}
\vspace*{2mm}
\begin{center}
\includegraphics[width=9.0cm]{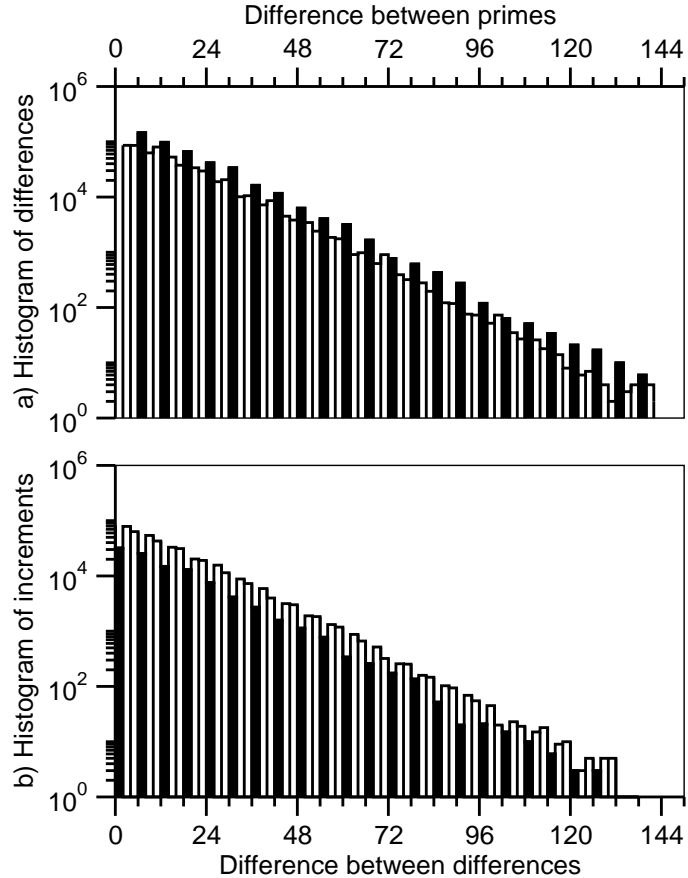}
\end{center}
\caption{a) Histogram of differences for
the first million of consecutive prime 
numbers. Filled black bars correspond to $6$, $12$, $18$,\ldots b) Histogram
of the increments (differences of differences). We have not plotted the
negative part of the $x$ axis in b) to simplify comparison between both
figures. In this work we are not concerned with the decreasing trend, 
but rather with the periodic oscillations observed.}
\label{fig1}
\end{figure}

Prime numbers have fascinated scientists
of all times,
and their history is closely related to the very history of
Mathematics. Recently, the interest in prime numbers has
received a new impulse because they have appeared in different
contexts ranging from Cryptology \cite{crypto} to
Biology \cite{bio1,bio2} or quantum chaos \cite{chaos,sakhr}, where the fine
structure in primes must reflect properties of very high Riemann zeros.
But, despite the huge advances in number theory, many
properties of the prime numbers are still unknown, and they appear
to us as a random collection of numbers without much structure. 
In the last few years, some numerical
investigations related with the statistical properties of the prime
number sequence \cite{wolf1,wolf2,stanley,dahmen} have revealed that,
apparently, some regularity actually exists in the
differences and increments (differences of differences) of consecutive
prime numbers. For instance, some oscillations are found in the
histogram of differences as we show in Fig.\ \ref{fig1}a. 
In that figure, one can see some
spikes located at positions $6$, $12$, $18$, and so on. A similar behavior was reported
by Kumar {\em et al.} \cite{stanley}, who showed that the histogram of increments
has also a similar periodicity; in this case there are some grooves at
increments given by $0$, $\pm 6$, $\pm 12$, etc (see Fig.\ \ref{fig1}b). 
These novel and promising findings
have been thought to provide new information about the underlying unpredictable
distribution of prime numbers and its potential applications.

The aim of this work is twofold: on the one hand we demonstrate 
that the apparent regularities 
observed by these authors do not reveal any structure in the sequence of primes, 
and that it is precisely a consequence of its randomness. On the other hand, we 
show that this randomness provides a new kind of predictable patterned 
behavior that we can characterize explicitly and compute analytically. 

The paper is organized as follows: first, we introduce a theoretical
framework to calculate the properties of consecutive differences of
prime numbers. After that, we check numerically the validity of the 
theory when finite sequences of primes are computed, and discuss
some of the main results obtained. 
Finally, conclusions are summarized.

\section{Theory}

Essentially, we will be dealing with the sequences obtained from the primes 
by subtracting them iteratively.
Our findings are based on two basic results: the first one is the fact that every 
prime $p>3$ is $p=\pm 1 (\mod 6)$ ({\em i.e.}, there exists an integer $N$ such that 
$p=6N\pm 1$); the second one, a theorem by Dirichlet which states that, for any 
pair of numbers $a$ and $q$ with no common divisors, there are infinitely many 
primes $p=a(\mod q)$. In addition, these primes are roughly equidistributed for 
each possible value of $a$ \cite{dirichlet}. Setting $q=6$ this means that 
$p=+1(\mod 6)$ and $p=-1(\mod 6)$ {\em equally likely}. In a probabilistic language, if 
$P_n(+1)$ denotes the fraction of primes $p\leq n$ which are $+1(\mod 6)$ 
(hence $P_n(-1)=1-P_n(+1)$), then
\begin{equation}
\lim_{n\rightarrow \infty} P_n(\pm 1)=\frac{1}{2},
\label{p+p-}
\end{equation}
Despite the apparent lack of structure that this result suggests, the numerical 
results concerning differences between primes cited above inspired us to search 
for regularities in the sea of randomness of the prime number sequence. 

Now, we build the sequence of differences of consecutive primes, $D^{(1)}$, the sequence 
of differences of consecutive differences,  $D^{(2)}$ and, in general, the m-differences, 
$D^{(m)}$, defined as differences of consecutive (m-1)-differences. For instance, 
taking the sequence of primes greater than $3$: $5$, $7$, $11$, $13$, $17$, $19$, 
$23$, $29$, $31$, $37$,\ldots, we find:
\begin{eqnarray}
D^{(1)}&=&2, 4, 2, 4, 2, 4, 6, 2, 6,\ldots,\label{d1}\\
D^{(2)}&=&2, -2, 2, -2, 2, 2, -4, 4,\ldots.\label{d2} 
\end{eqnarray}
Note that all numbers are even because all primes $p\geq 3$ are odd.

The structure of these m-differences becomes clearer when we write them modulo 6 
(this choice will be clarified below). Hereafter, we will term $d^{(m)}$
the integer between $-2$ and $2$ such that $d^{(m)}=D^{(m)}(\mod 6)$. 
As all the numbers in the sequences given by Eqs.\ (\ref{d1})-(\ref{d2})
are even, $d^{(m)}$ can only take the values $0$ and $\pm 2$. 
At this point, the reader will have noted that the origin of the periodicity 
seen in the histograms of differences is a consequence of the fact that
every prime number is $\pm 1(\mod 6)$. 
More explictly, we will find a $0$ in the $d^{(1)}$ sequence whenever two consecutive 
primes are both of the type $+1$ or both $-1$. Conversely, we will find $+2$ 
(respectively $-2$) only when consecutive primes are $-1$ and $+1$ ($+1$ and $-1$). 
Then, given equation (\ref{p+p-}), 
the probability of finding a $0$ in the sequence is twice that of finding $+2$ or $-2$. 
We want to remark that we have made the only extra assumption
that consecutive primes can be $\pm 1$ independently, 
that is, the correlations between consecutive primes are negligible.
Nevertheless, this is a hypothesis that we expect to hold for very large
sequences of primes (average correlations will disappear for long enough
sequences), but that we can not prove: this makes all our work a kind of
conjecture, or only an approximation. Note that, if we consider the prime
numbers to be $+1$ or $-1$ independently, taking the sequence in order
(consecutive primes) makes no
difference with considering differences between any pair of given primes,
even if they are not consecutive.

In the same way, we can rate the relative frequencies for the $d^{(2)}$ sequence, 
noting that the terms we are subtracting are not $\pm 1$, but $0$ and $\pm 2$. 
Therefore, the frequency 
of an outcome to be $0$ is $2/3$ times that of the frequencies to be $+2$ or $-2$. 
This corresponds to the grooves in the histogram of increments in 
Fig.\ \ref{fig1}b.

At this point, we can calculate iteratively the subsequent frequencies
for any m-difference. Nevertheless, we provide an exact formula to
calculate them at any order through the generating function of the $d^{(m)}$
sequences. This task can be achieved because the m-differences satisfy
some recurrence relations for a given piece of $m+1$ prime numbers. Let
$s_0\ldots s_m$ denote the value of those primes modulo $6$, then:
\begin{equation}
d^{(m)}(s_0,\ldots,s_m)= \sum_{k=0}^m \bi{m}{k} (-1)^k s_k.
\label{dm}
\end{equation}
So our search for $0$ (respectively $+2$ and $-2$) in the sequences requires
counting the number of solutions to the equation $d^{(m)}=0$ ($=+2$ and $=-2$).
For a probability distribution $p_t$ of a discrete variable $t$,
we can define
a generating function \cite{wilf}:
\begin{equation}
\label{eq:generating}
G(z)=\sum_t p_t z^t.
\end{equation}
The probability distribution in which we are interested is $P^t(m)$, which is
the probability that a given sequence of $m+1$
consecutive primes has an $m$-difference $d^{(m)}=t$, where $t$
can only take values $0$, $2$ and $-2$. It's associated generating function is:
\begin{equation}
\label{eq:generating_2}
G_m(z)=P^0(m)+P^2(m)z^2+P^{-2}(m)z^{-2}.
\end{equation}
We have made the assumption that the $s_i$ in a sequence $(s_0,\ldots,s_m)$ are
all independent, so the probabilities $P^t(m)$ are proportional to the number of
sequences such that $d^{(m)}(s_0,\ldots,s_m)=t$. Using this, we can write the
generating function in (\ref{eq:generating_2}) (forgetting the factor
$2^{-(m+1)}$, which accounts for the total number of different sequences
$(s_0,\ldots,s_m)$) as:
\begin{equation}
\label{eq:gen_delta}
G_m(z)=\sum_{s_0=\pm 1}\ldots \sum_{s_m=\pm 1}\left\{
z^{d^{(m)}}\delta(d^{(m)}-0)+z^{d^{(m)}}\delta(d^{(m)}-2)+
z^{d^{(m)}}\delta(d^{(m)}+2)\right\},
\end{equation}
which using $\delta(d^{(m)}-0)+\delta(d^{(m)}-2)+\delta(d^{(m)}+2)=1$
turns to be:
\begin{equation}
G_m(z)=\sum_{s_0=\pm 1}\ldots \sum_{s_m=\pm 1} z^{d^{(m)}}.
\end{equation}
This function has some interesting properties. For instance, $G_m(1)=2^{m+1}$
provides all the possible sequences of primes that we need to evaluate to
determine the relative frequencies of $0$ and $\pm 2$. Similarly,
$\frac{1}{6}\sum_{n=0}^5 G_m(e^{n\pi i/3})$ gives
us only those combinations of primes for which $d^{(m)}=0$. In other words,
\begin{figure}
\vspace*{2mm}
\begin{center}
\includegraphics[width=9.0cm]{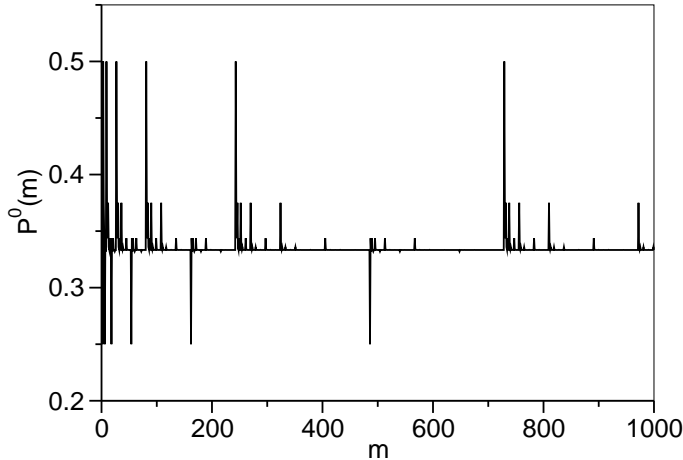}
\end{center}
\caption{Relative frequency of appearing $P_0(m)$ for the first $m=1000$ m-differences. 
Note that it displays a quasi-periodic behavior with $m$.}
\label{fig2}
\end{figure}
it helps us to calculate (performing the sums) the relative frequency of zeroes in the
m-difference sequence, $P^0(m)$ as 
\begin{equation}
P^0(m)=\frac{1}{3}\left(1+
\prod_{k=0}^m\cos\left[\frac{\pi}{3}\left(\begin{array}{c}m\\k\end{array} \right)\right]+
\prod_{k=0}^m\cos\left[\frac{2\pi}{3}\left(\begin{array}{c}m\\k\end{array} \right)\right]
\right).
\label{p0}
\end{equation}
Using a similar argument it can be shown that $P^{+2}(m)= P^{-2}(m)=(1- P^0(m))/2$.
Fig.\ \ref{fig2} shows $P^0(m)$ for the first $m=1000$ m-differences. 

\begin{figure}
\vspace*{2mm}
\begin{center}
\includegraphics[width=8.0cm]{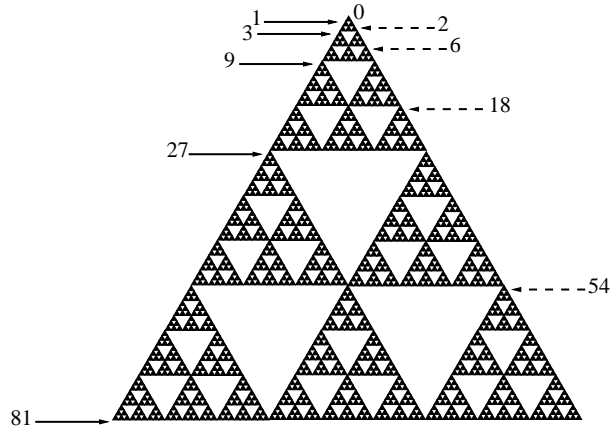}
\end{center}
\caption{Pascal's triangle modulo 3: we take the numerical triangle where the small 
black triangles are placed at those values that are non-zero ($\mod 3$). The resulting 
figure belongs to the family of Sierpinski's gaskets. Solid arrows show the empty rows 
(except the first and last element) that give maxima for $m$ in Fig.\ \ref{fig2}. Dashed arrows show 
empty rows except for the central element: they give the position of the minima.}
\label{fig3}
\end{figure}

The quasi-periodic behavior displayed by $P^0(m)$ is also remarkable. 
This behavior arises from the properties
of the binomial coefficients appearing in equation (\ref{p0}) and the periodicity
of the cosine. Thus,  a maximum of
$P^0(m)$ is found whenever all the elements in a row of Pascal's triangle 
(except the first
and the last, which are always $1$) are multiples of $3$. So,
the maxima are located at $m=3^n$ and the minima at $m=2\cdot3^n$,
where $n=0, 1, 2,\ldots$ These maxima and minima can be easily identified graphically
in Fig.\ \ref{fig3}, which represents a kind of Sierpinski's gasket obtained from the modulo $3$
Pascal's triangle. We want to stress that this is an analytical result, 
and that the correspondence
between the properties of $P^0(m)$ and Pascal's triangle can be shown to be
rigorously deduced from the properties of the binomial coefficients.

A connection can be found between our ideas and a classical number
theory
conjecture due to Hardy and Littlewood \cite{hardy} which states:

\begin{conj}[Hardy and Littlewood] Let $b_0,b_1,\ldots,b_m$ be $m$ distinct integers, and
$P(x;b_0,b_1,\ldots,b_m)$ the number of groups $n+b_0,n+b_1,\ldots,n+b_m$
between $1$ and $x$ consisting wholly of primes. Then
\begin{equation}
\label{eq:conj1}
\lim_{x\to\infty}P(x)\sim G(b_0,b_1,\ldots,b_m)
\int_2^x\frac{\textrm{d}x'}{(\ln x')^{m+1}},
\end{equation}
where
\begin{equation}
\label{eq:conj2}
G(b_0,b_1,\ldots,b_m)=\prod_{p\geq 2}\left(\left(\frac{p}{p-1}\right)^{m}
\frac{p-\nu}{p-1}\right),
\end{equation}
$\nu=\nu(p;b_0,b_1,\ldots,b_m)$ is the number of distinct residues of
$b_0,b_1,\ldots,b_m$ to modulus $p$, $p$ are the prime numbers and $n$ is a
natural number.
\footnote{As a curiosity, in the same reference Hardy and Littlewood
make another conjecture which states that
$\pi(x+y)-\pi(x)\leq\pi(y)$ for all $x$ and $y\geq 2$, where $\pi(x)$ is the
prime counting function. It was shown by Richards \cite{richards} that these
two conjectures are incompatible with each other. The one we reproduce in the
main text is generally believed to be true.}
\end{conj}
Formula (\ref{eq:conj2}) can be rewritten using that:
\begin{equation}
\label{eq:conjG}
G(b_0,b_1,\ldots,b_m)=C_m H(b_0,b_1,\ldots,b_m),
\end{equation}
where
\begin{equation}
\label{eq:Cm}
C_m=\prod_{p>(m+1)}\left(\left(\frac{p}{p-1}\right)^{m}
\frac{p-1-m}{p-1}\right),
\end{equation}
\begin{equation}
\label{eq:H}
H(b_0,b_1,\ldots,b_m)=\prod_{p\leq (m+1)}\left(\left(\frac{p}{p-1}\right)^{m}
\frac{p-\nu}{p-1}\right)\prod_{p|\Delta, p>(m+1)}\left(
\frac{p-\nu}{p-1-m}\right),
\end{equation}
and $\Delta$ is the product of the differences of the $b$'s. By $p|\Delta$ we
mean that $\Delta$ is divisible by $p$, $C_m$ are known as the Hardy-Littlewood
constants.
Written in this
form, all the dependence of $G$ on the actual members of the group of primes is
contained in $H$. But $H$ is a finite expression, and in principle can be
evaluated for all cases of interest. Moreover, as for fixed length
$m$ of a group of primes the other terms of $G$ are identical, the different
frequency of appearance of each type of groups of the same length depends only
on $H$. Note that this conjecture does not use that the primes
$n+b_0,n+b_1,\ldots,n+b_m$ have to be consecutive, in that sense, it has no
relationship with our results. But we have already said that our results should
hold also for non consecutive primes, due to the fact that they are only based in
considering independent occurrences of $+1$ and $-1$ primes. So we could try to
use the Hardy-Littlewood conjecture to compute the relative frequency
of different groups of primes. We are interested in what we have called
m-differences. A set of $n+b_0,n+b_1,\ldots,n+b_m$ differences between primes
defines a m-difference of order $m$ and value:
\begin{equation}
\label{bdiff}
D^{(m)}(b_0,\ldots,b_m)= \sum_{k=1}^m \bi{m}{k-1} (-1)^{k-1} b_k.
\end{equation}
Again, we denote $d^{(m)}=D^{(m)}(\mod 6)$.
The relative frequency of appearances of a given value $0$, $+2$ or $-2$ of 
$d^{(m)}$ for a constant $m$ is proportional to all the sets $(b_0,\ldots,b_m)$
that produce that value of $d^{(m)}$, and
from equations (\ref{eq:conj1}) and (\ref{eq:conjG})
the contribution of each set is
proportional to $H(b_0,\ldots,b_m)$.
We define then $\mathcal{H}^t(m)$ as the sum of all the $H(b_0,\ldots,b_m)$
such that $d^{(m)}(b_0,\ldots,b_m)=t$. This $\mathcal{H}^t(m)$ is
proportional to the number of
groups of consecutive primes $n+b_0,n+b_1,\ldots,n+b_m$ which have
$d^{(m)}(b_0,\ldots,b_m)=t$, and hence is proportional to the probability
$P^t(m)$ that a given m-difference has value $t(\mod 6)$.
\begin{figure}
\vspace*{2mm}
\begin{center}
\includegraphics[width=9.0cm]{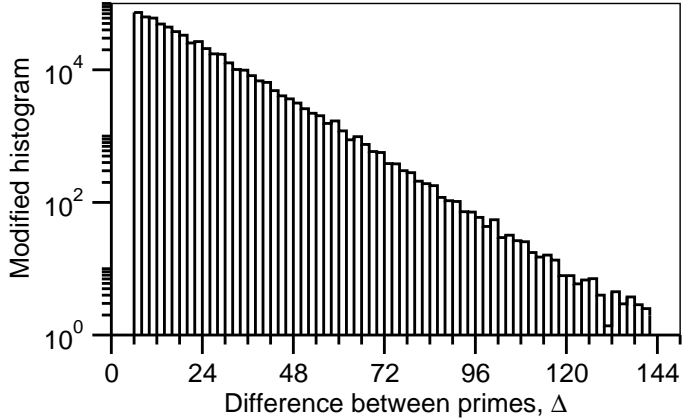}
\end{center}
\caption{\label{fig:prod} Histogram of differences
for the first million of consecutive prime 
numbers, divided by the factors $\prod_{p|\Delta}\frac{p-1}{p-2}$.}
\end{figure}
Using this, we can write:
\begin{equation}
\label{eq:pr0}
P^0(m)\propto \mathcal{H}^0(m)=
\sum_{\forall \textrm{set}{\bf b}\backslash d^{(m)}=0}
H(b_0,\ldots,b_m)
\end{equation}
\begin{equation}
\label{eq:pr2}
P^{+2}(m)\propto \mathcal{H}^{+2}(m)=
\sum_{\forall \textrm{set}{\bf b}\backslash d^{(m)}=+2}
H(b_0,\ldots,b_m)
\end{equation}
\begin{equation}
\label{eq:pr-2}
P^{-2}(m)\propto \mathcal{H}^{-2}(m)=
\sum_{\forall \textrm{set}{\bf b}\backslash d^{(m)}=-2}
H(b_0,\ldots,b_m)
\end{equation}
From our results, we make the following conjecture concerning the sums of
function $H$:
\begin{equation}
\label{eq:conjH2}
\mathcal{H}^{+2}(m)=\mathcal{H}^{-2}(m)\equiv\mathcal{H}^{2}(m),
\end{equation}
and from it we make a final conjecture that relates explicitly our formula
(\ref{p0}) with the Hardy-Littlewood theory:
\begin{equation}
\label{eq:finalconj}
P^0(m)=\frac{\mathcal{H}^{0}(m)}{\mathcal{H}^{0}(m)+2\mathcal{H}^{2}(m)}
\end{equation}
We have found no way to perform the infinite
sums in (\ref{eq:pr0}-\ref{eq:pr-2}), but we
find formula (\ref{eq:finalconj}) a beautiful relation and an interesting open
problem. This relation between our theory and the well established
Hardy-Littlewood theory shows that we can extract some information in our theory
directly from Hardy and Littlewood's developments, and therefore giving further
strength to our point. From equation (\ref{eq:H}) we can see, for instance, that
for $m=1$ (differences between consecutive primes), the probability of finding a
difference of value $\Delta$ is proportional to
$\prod_{p|\Delta}\frac{p-1}{p-2}$. Hence, dividing the histogram a) of Fig.\
\ref{fig1} by this factor, the periodicity disappears, as shown in Fig.\
\ref{fig:prod}. The case $m=1$ is specially simple because the factor $\nu$ is
trivially 1, and $\Delta$ is just the value of the difference between primes.
For higher $m$-differences the problem is harder, as we can no longer set $\nu$
and $\Delta$ just from the values of $m$ and the $m$-difference we are
interested in: in this case, for each $\Delta$ it is necessary to find all the
groups $(b_0,\ldots,b_m)$ from which it can be obtained, and compute the whole
factor $H(b_0,\ldots,b_m)$ associated with them.
Our theory establishes a connection between the {\em local}
properties (in the sense that
they only apply to single groups $(b_0,\ldots,b_m)$) in the Hardy-Littlewood
theory and {\em global} properties we describe, namely the $m$-differences,
which depend on {\em all} the possible groups $(b_0,\ldots,b_m)$. It gives,
through equation (\ref{p0}), a prediction that is easily computed for any
$m$, while from Hardy and Littlewood's results it is very difficult to say
anything beyond $m=1$.\\

\section{Numerical Results and Discussion}

All the theoretical results presented in Sec. II stand if we consider the
infinite sequence of primes.
Now we consider the situation in which we pick a finite sequence of primes. In such a 
case, we expect some deviations from the results presented above due to, for instance,
the transient behavior known as Chebyshev's bias. Chebyshev noted that at the beginning of
the sequence there are more primes of type $-1$ than of type $+1$. Moreover,
Bays and Hudson \cite{bays} proved that the first time when $P_n(+1)>P_n(-1)$ occurs for
$n=608,981,813,029$. This huge number is
of the order of magnitude of $10^{13}$,
bigger than $2^{39}$ and higher than numbers used for usual calculations
without using specialized software.
Nevertheless, this is a misleading result because the
relative error of considering that $P_n(+1)= P_n(-1)=1/2$ is about $1.6\%$
for the first $1,000$ primes, about $0.2\%$ for the first $10,000$ and just $0.08\%$
for the first $100,000$. This relative error continues to decrease for increasing number
of primes, so any finite sequence will reproduce our predictions with enough accuracy.
Moreover, from the work of Littlewood \cite{littlewood} it is now well known that the
inequality $P_n(-1)>P_n(+1)$ is reversed for infinitely many integers.

\begin{figure}
\vspace*{2mm}
\begin{center}
\includegraphics[width=9.0cm]{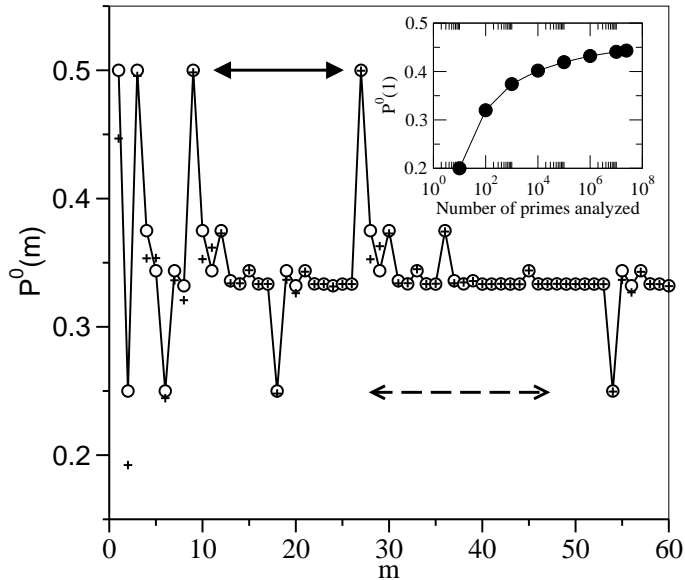}
\end{center}
\caption{Relative frequency of appearing $P_0(m)$ for the first $m=60$ m-differences. 
Circles stand for the theoretical value and pluses for the value obtained computing 
such m-differences for all the primes lower than $2^{31}$. The arrows show more clearly 
the position of the maxima and minima exposed at Fig.\ \ref{fig3}. Inset: numerical values of
$P_0(1)$ as a function of the length of the sequence analyzed. Although slow,
the convergence to the exact value ($0.5$) is monotonous.}
\label{fig4}
\end{figure}

In Fig.\ \ref{fig4} we show both theory and numerics. As we anticipated, there are some slight
deviations but just at a small number of points. We have checked that these deviations
reduce monotonously as we take larger sequences, due to the vanishing of Chebyshev's bias
as we increase the sequence size (see inset in Fig.\ \ref{fig4}). 
One criticism may be that the error is significantly large
in some cases, but note that this error is not large enough to hide the periodicity in the
sequences. For instance, for one of the worst cases, $m=1$, we predicted
$P^0(1)=2P^{\pm 2}(1)$ but, actually, we have found numerically $P^0(1)=1.62P^{\pm 2}(1)$.
Thus, the periodicity modulo $6$ is still transparent in spite of the fact 
that its strength deviates
from the predicted one.

The reason why we use $6$ is that it is the first factor giving non-trivial
information about the prime numbers. The first number to give us information
about them is 2: it says that the rest of the primes cannot be multiples of 2.
The same applies for 3. 4 gives no new information, as it is $4=2^2$, so it only
says that prime numbers cannot be multiple of 2. 5, of course, excludes its own
multiples. 6 is the first product of two different primes, and hence, saying
that the prime numbers are of the form $6n\pm 1$ says that they cannot be
multiples of 2 nor multiples of 3. From this simple
observation, the fact that $m$-differences have to be $0(\mod 6)$ or
$\pm 2(\mod 6)$ is readily derived. If we tried to repeat the same formalism
with 4, we should see that all the primes are $4n\pm 1$, which is the same as
saying that they are $1(\mod 2)$, that is, odd numbers. $m$-differences would be
$0(\mod 4)$ or $\pm 2(\mod 4)$. But it has no sense considering this difference
of sign between $+2(\mod 4)$ and $-2(\mod 4)$, as both things are exactly the
same. So we would only be able to talk about $0(\mod 4)$ and $2(\mod 4)$, which
means that both of them are equiprobable, and so no periodicity with period 4
should be observed, as in fact happens.
Our results can be generalized taking other divisors $q$ instead of $6$.
As 6 is $2\times 3$ and thus has non-trivial information about the sequence
of prime numbers, the same happens with all the different products of primes.
That is, our formalism can be extended to $2\times 5=10, 2\times 7=14,
2\times 11=22, 2\times 13=26, 2\times 3\times 5=30$, etc., predicting secondary
periodicities in the $m$-differences histograms. This periodicities have already
been observed in references \cite{wolf1,wolf2}, and are explained in the
framework of our theory.

Moreover, we may also consider the case in which the differences are
evaluated between non-consecutive primes provided that they are picked at random. 
The results will remain the same because all the differences are calculated modulo $6$. 
For instance, a chaotic system with energy levels proportional to prime numbers will
display oscillations in the spectrum of the emitted light through radiative
transitions, because the frequency of the emitted light is proportional to
the difference between energy levels (consecutive or not).

Another observation that supports our findings is the jumping champion
phenomenon: at the beginning of the prime numbers sequence, the most often
occurring 1-difference is 6. From $10^{35}$ it is 30, and
there is evidence that from
around $10^{450}$ it is 210, etc. In reference \cite{odlyzko} this phenomenon,
also related with the Hardy-Littlewood conjecture, is studied in detail. Table 3
in this reference shows histograms of gaps between consecutive primes
\footnote{In fact, {\em probable} primes.} for primes next to $10^{30}$,
$10^{40}$, $10^{400}$ and $10^{450}$. Although statistics is not good enough in the
biggest cases, in all of them the periodicity of period 6 is clearly displayed,
giving a numerical evidence of our findings far beyond our own numerical
calculations. Note also that all the known jumping champions are multiples of 6,
as should be expected.

Finally, we can reformulate all the presented findings referring to the fact that the 
random structure of the $\pm 1$ primes resembles the high temperature (disordered) 
phase of spin systems \cite{huang}. This analogy can be cast explicitly by means of 
the Hamiltonian
\begin{equation}
{\mathcal H}\equiv d^{(m)}(s_0,\ldots,s_m)= \sum_{k=0}^m \bi{m}{k}(-1)^k s_k.
\end{equation}
Then, the generating function $G_m(z)$ is equal to the partition function of ${\mathcal H}$,
${\mathcal Z}$,
if we identify the temperature as $k_BT=-1/\log z$. Note that this
would correspond to a spin system defined on a one dimensional lattice
without interactions between spins and subjected to an 
applied external field 
\begin{equation}
B_k=\bi{m}{k}(-1)^k.
\end{equation}
Likewise, we could take some advantage of the collected knowledge in spin systems
to gain some insight into the properties of the products between primes. These
products are the basis of some encrypting systems~\cite{crypto}.
For instance, let us consider an interacting Hamiltonian of the form 
\begin{equation}
{\mathcal H}=\sum_{i,j} J_{ij}s_is_j.
\end{equation}
where the coupling constants $J_{ij}$ would depend on the recurrence relations
between sequences of products. We expect this kind of approach to hold because
the product of two primes will also be of the form $\pm 1(\mod 6)$.
     
\section{Conclusions}

We have studied the general behavior of consecutive differences between primes,
both theoretically and numerically. Our theoretical predictions
are based on a theorem by Dirichlet and on a hypothesis that neglects
a kind of correlations between prime numbers.
In principle they are
only valid for the whole sequence of primes,
but we find that they
are also accurate for finite
sequences. The deviations found are due to the transient behavior known as
Chebyshev's bias.
Furthermore, the theory is still valid if the differences are computed between
non-consecutive primes, and Chebyshev's bias will be less pronounced.

Our main conclusion is that the main feature of the sequence of prime numbers, namely,
its randomness, hides an underlying behavior arising when successive differences are 
computed.
Thus, new and interesting phenomena can be derived and novel applications can be 
settled. For instance, the length of the periodic orbits in quantum chaos is related to 
the zeroes of Riemann's function and these with the prime numbers~\cite{sakhr}.
So, these new 
findings can be of interest in the study of their statistical properties and engage
with random matrix theory and the most outstanding problem in number theory:
Riemann's hypothesis~\cite{chaos}. We have related our theory to
established number theory through the Hardy-Littlewood conjecture.
We have also found a connection between our theory and the statistical
mechanics of spin systems.

To conclude, we mention another interesting example in which the differences
between primes are crucial, which is related to the fact that the life
cycles of different animal species are precisely prime numbers. 
In this case, the life or death of the species depends on this property~\cite{bio1}.

\begin{ack}
We are indebted to Jos\'e A.\ Cuesta for valuable discussions and critical reading 
of the manuscript, and with Sara Cuenda for her collaboration in preparing 
the figures. This work has been supported by the Ministerio de Ciencia y Tecnolog\'{\i}a 
of Spain through grants BFM2003-07749-C05-01 and BFM2003-07749-C05-05.
\end{ack}


\begin{thebibliography}{00}

\bibitem{crypto}
M. Stallings,
Cryptography and network security: principles and practice,
(Prentice Hall, New Jersey, 1999)
\bibitem{bio1}
E. Goles, O. Schulz and M. Markus,
{\em Complexity\/} {\bf 6} (2001) 33.
\bibitem{bio2}
J. Toh\'a and M. A. Soto,
{\em Medical Hypotheses\/} {\bf 53(4)} (1999) 361.
\bibitem{chaos}
M. V. Berry,
{\em Inst. Phys. Conf. Ser. No. 133\/} (1993).
\bibitem{sakhr}
J. Sakhr, R. K. Bhaduri and B. P. van Zyl,
{\em Phys.\ Rev.\ E\/} {\bf 68} (2003) 026206.
\bibitem{stanley}
P. Kumar, P. C. Ivanov and H. E. Stanley,
{\em cond-mat/0303110\/} (2003).
\bibitem{wolf1}
M. Wolf,
{\em Proc. of the 8th Joint EPS-APS Int. Conf. Physics Computing'96\/},
P. Borcherds et al, eds.
(Krak\"ow, 1996) s. 361.
\bibitem{wolf2}
M. Wolf,
{\em Physica A\/} {\bf 274} (1999) 149.
\bibitem{dahmen}
S. R. Dahmen, S. D. Prado and T. Stuermer-Daitx,
{\em Physica A\/} {\bf 296} (2001) 523.
\bibitem{dirichlet}
L. Dirichlet,
{\em K\"onig. Preuss. Akad.\/} {\bf 34} (1837) 45.
Reprinted in Dirichlets Werke, vol 1,
(Reimer, Berlin and Chelsea, Bronx (NY), 1889--97 and 1969).
\bibitem{wilf}
H. S. Wilf,
Generating Functionalogy,
2nd ed. Academic Press, (New York, 1994).
\bibitem{hardy}
G. H. Hardy and J. E. Littlewood,
{\em Acta Math.\/} {\bf 44} (1923) 1.
\bibitem{richards}
I. Richards,
{\em Bull. Amer. Math. Soc.\/} {\bf 80} (1974) 419.
\bibitem{bays}
C. Bays and R. H. Hudson,
{\em Math. Comp.\/} {\bf 32} (1978) 571.
\bibitem{littlewood}
J. E. Littlewood,
{\em Comptes Rendus} {\bf 158} (1914) 1869.
\bibitem{odlyzko}
A. Odlyzko, M. Rubinstein and M. Wolf,
{\em Exp.\ Math.\/} {\bf 8} (1999) 107.
\bibitem{huang}
K. Huang,
Statistical Mechanics,
2nd ed. John Willey \& Sons, (New York, 1987).
\end{thebibliography}
\end{document}